\documentclass[pre,reprint,showpacs,floatfix,amsmath,amssymb]{revtex4-1}

\usepackage{amsmath, amssymb, amsfonts, amsthm}
\usepackage{verbatim}
\usepackage{hyperref}
\usepackage{graphicx,color}

\newcommand{\figOne}{
\begin{figure}[t]
    \centering
        \includegraphics[width=2.5in]{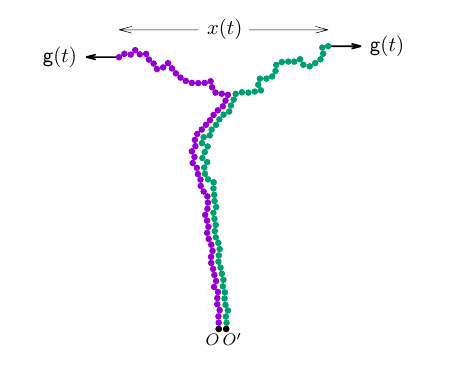}
    
	\caption {Schematic diagram of a dsDNA. One end of the DNA is
	anchored at the origin ($O$ and $O^{\prime}$) and the strands on the
	free end are subjected to a time-dependent periodic force
	$\mathrm{g}(t)$ with frequency $\omega$ and amplitude $G$.
	\label{fig:1}}

\end{figure}
}

\newcommand{\figTwo}{
\begin{figure*}[t]
\centering
 \includegraphics[width=0.85\linewidth]{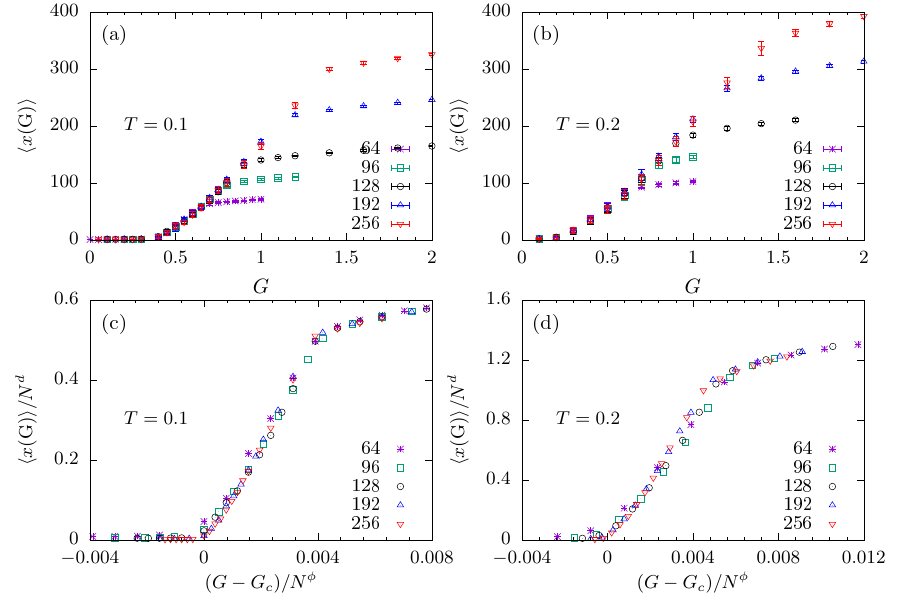}

	\caption{Average separation $\langle x (\mathrm{G}) \rangle$
	between the free strands of the DNA as a function of pulling force
	$G$ for various chain lengths $N=64$, 96, 128, 192, and 256 at (a)
	$T=0.1$ and (b) $T=0.2$.  Plot of scaled separation $\langle x(G)
	\rangle/N^d$ vs $(G-G_c)/N^{\phi}$ showing a nice collapse for (c)
	$T=0.1$ with exponents $d=1.10 \pm 0.05$, $\phi = 1.00 \pm 0.05$ and
	critical force $G_c (T=0.1) = 0.45 \pm 0.05$ and (d) for $T=0.2$
	with exponents $d=1.05\pm 0.05$, $\phi = 1.00 \pm 0.05$, and critical
	force $G_c (T=0.2) = 0.25 \pm 0.05$.} \label{fig:2} 

\end{figure*}
}

\newcommand{\figThree}{
\begin{figure}[t]
\centering
\includegraphics[width=0.95\linewidth]{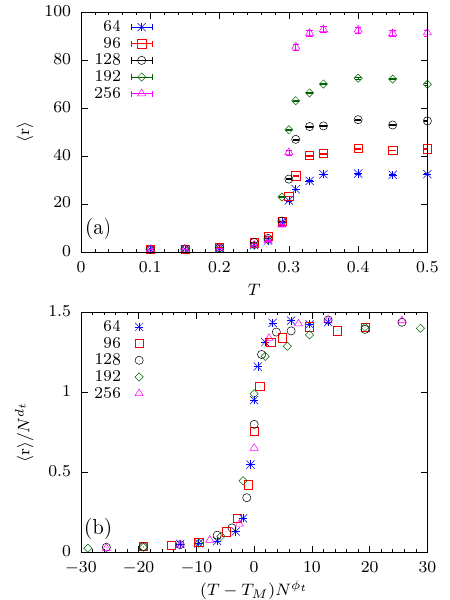}

	\caption{(a) Average separation $\langle \mathrm{r} \rangle$
	between the end monomers of two strands of the DNA as a function of
	temperature $T$ for various chain lengths. (b) Scaled separation
	$\langle \mathrm{r} \rangle/N^{d_t}$ vs $(T-T_M)N^{\phi_t}$ showing
	a nice collapse for exponents $d_t=0.75 \pm 0.05$, $\phi_t = 1.00 \pm
	0.05$, and the melting temperature $T_M = 0.30 \pm 0.01$.}
	\label{fig:3} 

\end{figure}
}

\newcommand{\figFour}{
\begin{figure}[t]
\centering
\includegraphics[width=0.95\linewidth]{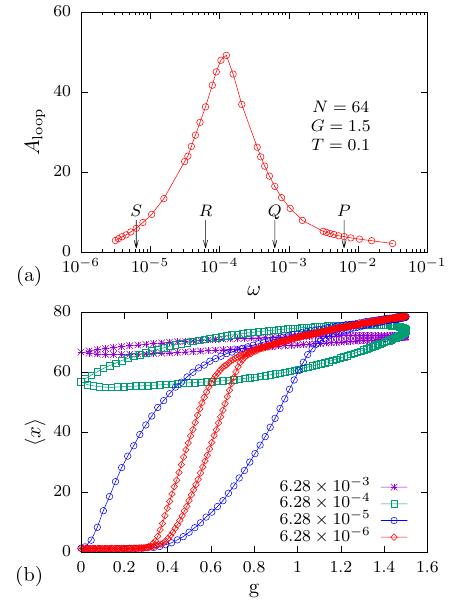}

\caption{(a) Area of hysteresis loop $A_{\mathrm{loop}}$ as a function
	of frequency $\omega$, in a semilog scale, for the DNA of length
	$N=64$ and force amplitude $G=1.5$ at $T=0.1$. (b) Average extension
	$\langle x \rangle$ as a function of force $\mathrm{g}$ as various
	frequencies indicated in (a) by arrows. The line joining the points
	in these plots is just a guide for the eye.} \label{fig:4} 

\end{figure}
}

\newcommand{\figFive}{
\begin{figure}[t]
\centering
\includegraphics[width=0.95\linewidth]{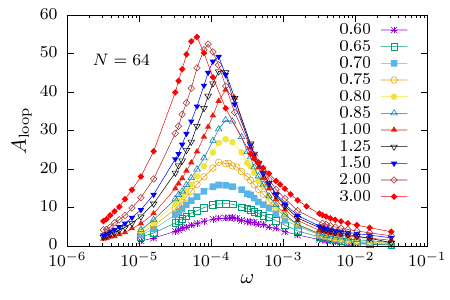}

\caption{Area of hysteresis loop $A_{\mathrm{loop}}$ as a function of
	frequency $\omega$, in a semilog scale, for the DNA of length
	$N=64$ at various force amplitudes $G$. The line joining the points
	in these plots is just a guide for the eye.} \label{fig:5} 

\end{figure}
}

\newcommand{\figSix}{
\begin{figure*}[t]
\centering
\includegraphics[width=0.95\linewidth]{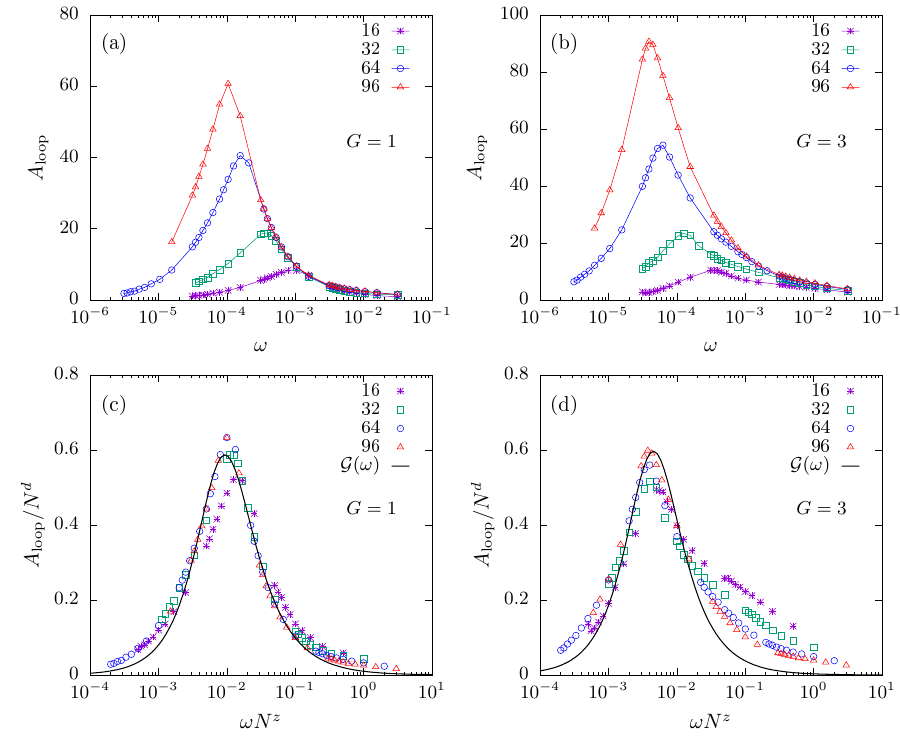}

\caption{Area of hysteresis loop $A_{\mathrm{loop}}$ as a function of
	frequency $\omega$, in a semilog scale, for DNA of lengths $N=16$,
	32, 64, and 96 for force amplitudes (a) $G=1$ and (b) $G=3$. Scaled
	loop area, $A_{\mathrm{loop}}/N^{d}$, plotted against $\omega N^{z}$
	showing a nice collapse for exponents (c) $d=1.00 \pm 0.05$ and
	$z=1.00\pm 0.05$ for $G=1$, and (d) $d=1.15 \pm 0.05$ and $z=1.00\pm
	0.05$ for $G=3$. The line joining the points in these plots is just a
	guide for the eye.} \label{fig:6} 

\end{figure*}
}

\newcommand{\figSeven}{
\begin{figure*}[t]
\centering
\includegraphics[width=0.95\linewidth]{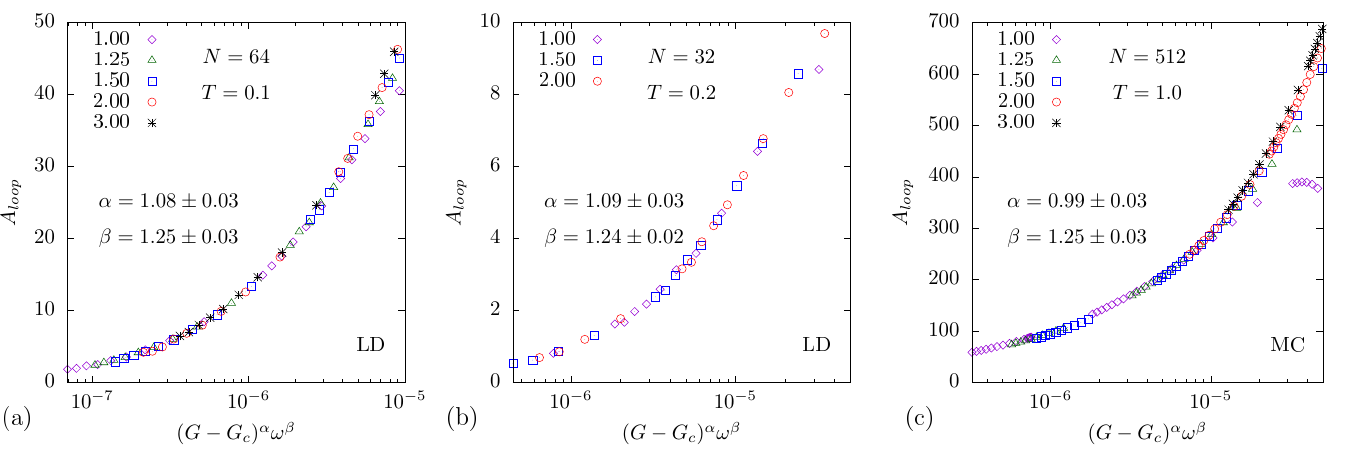}

	\caption{$A_{\mathrm{loop}}$ as a function of
	$(G-G_c)^{\alpha}\omega^{\beta}$ for various force amplitudes
	obtained (a) using LD simulations for the DNA of length $N=64$ at
	$T=0.1$ with $G_c(T=0.1) = 0.45 \pm 0.05$ with exponents
	$\alpha=1.08\pm0.03$ and $\beta=1.25\pm0.03$, (b) using LD
	simulations for the DNA of length $N=32$ at $T=0.2$ with $G_c(T=0.2)
	= 0.25 \pm 0.05$ with exponents $\alpha=1.09\pm0.03$ and
	$\beta=1.24\pm0.02$, (c) using MC simulations of the DSAW model of
	DNA of length $N=512$ at $T=1$ with $G_c (T=1) = 0.678$ with
	exponents $\alpha=0.99\pm0.03$ and $\beta=1.25\pm0.03$, as obtained
	in Ref.~\cite{kapri_unzipping_2014}. } \label{fig:7}

\end{figure*}
}

\begin{document}

\title{Hysteresis loop area scaling exponents in DNA unzipping by a
periodic force: A Langevin dynamics simulation study}

\author{Rajeev Kapri}
\email{rkapri@iisermohali.ac.in}

\affiliation{Department of Physical Sciences, Indian Institute of
Science Education and Research Mohali, Sector 81, Knowledge City, S. A.
S. Nagar, Manauli PO 140306, India.}

\date{\today}

\begin{abstract}
    
	Using Langevin dynamics simulations, we study the hysteresis in
	unzipping of longer double-stranded DNA chains whose ends are
	subjected to a time-dependent periodic force with frequency $\omega$
	and amplitude $G$ keeping the other end fixed. We find that the area
	of the hysteresis loop, $A_{loop}$, scales as $1/\omega$ at higher
	frequencies, whereas it scales as $(G-G_c)^{\alpha}\omega^{\beta}$
	with exponents $\alpha=1$ and $\beta=1.25$ in the low-frequency
	regime. These values are same as the exponents obtained in Monte
	Carlo simulation studies of a directed self-avoiding walk model of a
	homopolymer DNA [R. Kapri, \href {\doibase
	10.1103/PhysRevE.86.041906}{Phys. Rev. E {\bf 90}, 062719 (2014)}],
	and the block copolymer DNA [R. K. Yadav and R. Kapri, \href
	{\doibase 10.1103/PhysRevE.103.012413}{Phys. Rev. E {\bf103}, 012413
	(2021)} ] on a square lattice, and differs from the values reported
	earlier using Langevin dynamics simulation studies on a much shorter
	DNA hairpins.

\end{abstract}

\maketitle

\section{Introduction}

The unzipping of a double stranded DNA (dsDNA) by a mechanical force,
which is an initial step in biological processes like DNA replication
and RNA transcription~\cite{watson_molecular_2003}, has been studied
over two decades both
theoretically~\cite{bhattacharjee_unzipping_2000,lubensky_pulling_2000,sebastian_pulling_2000,marenduzzo_phase_2001,marenduzzo_dynamical_2002,kapri_complete_2004,kumar_biomolecules_2010}
and experimentally using single-molecule manipulation
techniques~\cite{bockelmann_unzipping_2002,danilowicz_dna_2003,danilowicz_measurement_2004,ritort_single-molecule_2006,hatch_measurements_2007}.
The two strands of a dsDNA, whose ends are exerted by a pulling force,
is found to unzip to two single strands if the force exceeds a critical
value. The unzipping transition, which is now well established, is a
first-order phase transition. If the externally applied force is
oscillatory in nature, then it unzips and rezips the two strands of the DNA
in each cycle, and the force-extension isotherm shows a hysteresis.
There have been many studies of hysteresis in unbinding and rebinding of
biomolecules in recent years because it reveals important information
about the kinetics of conformational transformations, the potential
energy landscape, and controlling the folding pathway of a single
molecule and in force sensor
studies~\cite{hatch_measurements_2007,friddle_near-equilibrium_2008,tshiprut_exploring_2009,li_real-time_2007,yasunaga_quantifying_2019}.

In recent years, the behavior of a dsDNA under a periodic force with
frequency $\omega$ and amplitude $G$ has been studied by using Brownian
dynamics (BD) or Langevin dynamics (LD) simulation of an off-lattice
coarse-grained model for short chains which are limited to a maximum
number of $N=16$ base pairs and 32
monomers~\cite{kumar_statistical_2013,mishra_dynamical_2013,mishra_scaling_2013,kumar_periodically_2016,pal_dna_2018},
and by using Monte Carlo (MC) simulations of DNA chains having 1024
monomers with $N=512$ base pairs on a ($D=1+1$)-dimensional square
lattice~\cite{kapri_hysteresis_2012,
kapri_unzipping_2014,kalyan_unzipping_2019,yadav_unzipping_2021}. Both
LD and MC simulation studies show the existence of a dynamical phase
transition, where the DNA can be taken from the zipped state to an
unzipped state via a new dynamical state. The area of the hysteresis
loop, $A_{loop}$, which represents the energy dissipated in the system,
is found to scale as $1/\omega$ at higher frequencies, whereas at low
frequencies, $A_{\mathrm{loop}}$ scales as $G^{\alpha} \omega^{\beta}$,
or $(G-G_c)^{\alpha} \omega^{\beta}$, where $G_c$ is the critical force
needed to unzip the DNA for the static force case. The values of the
exponents $\alpha$ and $\beta$ obtained in BD/LD and MC simulations are,
however, different. In BD/LD simulation studies on shorter DNA
hairpins~\cite{kumar_statistical_2013,mishra_dynamical_2013,mishra_scaling_2013,kumar_periodically_2016,pal_dna_2018},
a chain having $2N$ monomers, whose first $N$ monomers are complementary
to the rest half. The monomers of the chain are chosen in such a manner
that the $i$th monomer from the anchored end can bind only with the
$(N-i)$th monomer of the chain, thus mimicking a base pair of the DNA.
One end of the chain is anchored at the origin and an external time
dependent periodic force $g(t)$ is applied on the free end along $x$
direction and its distance from the origin, $x(t)$, is monitored. In MC
simulation
studies~\cite{kapri_hysteresis_2012,kapri_unzipping_2014,kalyan_unzipping_2019,yadav_unzipping_2021},
the strands of DNA are represented by two directed self-avoiding walks
(DSAWs), which do not cross each other, on a ($D=1+1$)-dimensional
square lattice. Whenever the $i$th monomers of  walks (mimicking
complementary bases) are unit distance apart, there is a gain in energy
(base pairing). Two strands of the DNA at one end are always kept fixed
at origins $O$ and $O^{\prime}$ and the other end monomers are subjected
to a time-dependent periodic force $g(t)$ and the separation, $x(t)$,
between them is monitored. In both BD/LD and MC simulation studies, the
average force-distance isotherms obtained from the time series show
hysteresis loop whose area is studied as a function of $G$ and $\omega$.
Initial BD/LD simulation studies reported exponent values $\alpha=
\beta=1/2$~\cite{kumar_statistical_2013,mishra_dynamical_2013,mishra_scaling_2013}.
These were later modified to $\alpha=0.33$ and $\beta =
1/2$~\cite{kumar_periodically_2016}. However, a different set of
exponents, $\alpha=1$ and $\beta=1.25$, were obtained for longer
homopolymer DNA chains in MC simulation
studies~\cite{kapri_unzipping_2014}. Very recently, the DSAW model has
been extended to study the unzipping of a block copolymer DNA subjected
to a periodic force, and the same set of exponents ($\alpha=1$ and
$\beta=1.25$) were obtained~\cite{yadav_unzipping_2021}. This inspired
us to perform LD simulation studies on a longer DNA chains and
investigate the true values of the loop area exponents at low-frequency
regime. Our hypothesis is that the different set of exponents obtained
in earlier BD/LD studies, as compared to MC studies, are due to the
presence of strong finite-size effects because of shorter chain lengths
used in those studies.
 
In this paper, we study the unzipping transition of a dsDNA subjected to
a periodic force using LD simulations in two dimensions (2D) and compare
our results to a well-established DSAW model of a DNA on a $D=1+1$
square lattice. The later model has been studied extensively, for over
two decades, using the generating function, exact transfer matrix, and
Monte Carlo
techniques~\cite{marenduzzo_phase_2001,marenduzzo_dynamical_2002,kapri_complete_2004,kapri_unzipping_2006,kapri_randomly_2007,kapri_manipulating_2008,kapri_hysteresis_2012,kapri_unzipping_2014,kapri_unzipping_2014,kalyan_unzipping_2019,yadav_unzipping_2021}.
In unzipping transition, the average of the relative distance
$\mathsf{r} = |\mathbf{r_2}(N) - \mathbf{r_1}(N)|$ between the end
monomers of the two strands of the DNA (the order parameter), which is
the conjugate variable to an externally applied force, is always along
the direction of the force. The fluctuations in the transverse
directions are so small that they can be safely neglected.  In the
absence of any external pulling force, the two strands of the DNA can
also be denatured thermally, purely due to the competition between the
entropy and the energy, at a temperature $T_M$ known as the melting
temperature of DNA. Unlike thermal melting, which depends on both the
model and dimension used, the force-induced transition at  $T < T_M$ was
found to be independent of both the model and the dimension. The values
of the critical force and the melting temperature are, however, model
dependent~\cite{marenduzzo_phase_2001,marenduzzo_dynamical_2002}. The
length of the DNA simulated in this paper for the periodic case (up to
192 monomers with $N=96$ base pairs) are six times longer than the chain
lengths used in earlier BD/LD simulation
studies~\cite{kumar_statistical_2013,mishra_dynamical_2013,mishra_scaling_2013,kumar_periodically_2016,pal_dna_2018}.
We first consider the static force case and confirm that the model
considered in this paper indeed show first-order phase transition same
as DSAW model. We obtain the value of the critical force,
$G_c(T)$, needed to unzip the dsDNA at two different temperatures and
also the melting temperature $T_M$ for the model. Next, we consider the
periodic force case, where the force-distance isotherms show hysteresis
loop whose area, $A_{\mathrm{loop}}$, behaves nonmonotonically with the
frequency. We observe that the loop area scales as $A_{\mathrm{loop}}
\sim 1/\omega$ in the higher-frequency regime, whereas it scales as
$A_{\mathrm{loop}} \sim (G-G_c)^{\alpha}\omega^{\beta}$ with exponent
values $\alpha \approx 1$ and $\beta \approx 1.25$ in the
lower-frequency regime. These exponents are similar to the exponents
obtained in earlier MC simulation studies on longer chain
lengths~\cite{kapri_unzipping_2014,yadav_unzipping_2021}.

The paper is organized as follows: In Sec.~\ref{sec:model}, we define
the model simulated in this paper. The results are discussed in
Sec.~\ref{sec:results} and summarized in Sec.~\ref{sec:summary}.

\section{Model} \label{sec:model}

\figOne

We model the strands of a dsDNA by beads and springs in two dimensions
(see Fig.~\ref{fig:1}). The beads of the polymer experience an excluded
volume interaction modeled by the Weeks-Chandler-Andersen potential of 
the form
\begin{equation}\label{eq:LJmm}
    U_{\mathrm{bead}}(r) = \begin{cases}
        4 \varepsilon \left[ \left( \frac{\sigma}{r} \right)^{12} - 
		\left( \frac{ \sigma}{r} \right)^{6} \right] + \varepsilon  
		& \text{for} \ r \le r_{\mathrm{min}} \cr
        0  & \text{for} \ r > r_{\mathrm{min}}
    \end{cases}
\end{equation}
where, $\varepsilon$ is the strength of the potential.  The cutoff
distance, $r_{\mathrm{min}} = 2^{1/6} \sigma$, is set at the minimum of
the potential. The consecutive monomers of strands are connected by the
finitely extensible nonlinear elastic (FENE)
potential~\cite{grest_molecular_1986} of the form
\begin{equation}\label{eq:FENE}
    U_{\mathrm{FENE}} (r) = - \frac{1}{2} k R_{0}^{2} \ln \left( 1 -
	\frac{r^2}{R_0^2}\right),
\end{equation}
where $k$ and $R_0$ are the spring constant and the maximum allowed
distance between the consecutive monomers, respectively. The
complementary monomers of the DNA (i.e., $i$th monomers of both the
strands) interacts with each other via standard LJ potential:
\begin{equation}\label{eq:LJbp}
    U_{\mathrm{bp}}(r) = \begin{cases}
	    4 \varepsilon_{\mathrm{p}} \left[ \left( \frac{\sigma}{r}
		\right)^{12} - \left( \frac{ \sigma}{r} \right)^{6} \right] &
		\text{for} \ r \le r_{\mathrm{c}} \cr
	    0  & \text{for} \ r > r_{\mathrm{c}},
    \end{cases}
\end{equation}
where $\varepsilon_{\mathrm{p}}$ denotes the base pair interaction
strength and $r_{\mathrm{c}} = 2.5 \sigma$ is the cutoff distance. 

The strands at one end of the DNA are anchored at $O$ and $O^{\prime}$,
which are $1.12\sigma$ distance apart, and the strands at the free end
are subjected to a time-dependent periodic force
\begin{equation}
    \mathbf{g}(t) = G | \sin (\omega t) |,
\end{equation}
where $G$ is the amplitude and $\omega$ is the angular frequency of the
oscillating force.

To integrate the equation of motion for the monomers of the chain we use
LD algorithm with velocity-Verlet update. The equation of motion for a
monomer is given by
\begin{equation}
    m \ddot{\boldsymbol r}_i = - {\boldsymbol \nabla} U_{i} +
	{\mathbf{g}} - \zeta {\boldsymbol v}_i + {\boldsymbol \eta}_i,
\end{equation}
where $m$ is the monomer mass, $ U_i = U_{\mathrm{bead}} +
U_{\mathrm{FENE}} +  U_{\mathrm{bp}}$ is the total potential experienced
by $i$th monomer, $\zeta$ is the friction coefficient, ${\boldsymbol
v}_i$ is the monomer's velocity, and ${\boldsymbol \eta}_i$ is the
random force satisfying the fluctuation-dissipation theorem $\langle {
	\eta}_i(t){\eta}_j(t^{\prime}) \rangle = 2 \zeta k_{B} T \delta_{ij}
\delta(t - t^{\prime})$. The unit of energy, length, and mass are set by
$\varepsilon$, $\sigma$, and $m$, respectively, which sets the unit of
time as $\tau = \sqrt{m \sigma^2 / \varepsilon}$. In these reduced
units, we choose $\zeta = 1.0$, $\varepsilon_{\mathrm{p}} =
\varepsilon$, $k = 30 \varepsilon/\sigma$, $R_0 = 1.5\sigma$, and $k_B T
= 0.1 \varepsilon$. The force is measured in units of
$\sigma/\varepsilon$. A time step of $\Delta t = 0.005$ is used in all
simulation runs. The simulations are done using LAMMPS
software~\cite{plimpton_LAMMPS_1995}.

The distance between the end monomers of the two strands is monitored as
a function of time, $x(t)$, for various force amplitudes $G$ and
frequency $\omega$. Due to the periodic nature of the applied force, the
extension $x(\mathrm{g})$ as a function of force $\mathrm{g}$ can be
obtained from the time series $x(t)$. This is then averaged over $1000$
cycles to obtain the average extension, $\langle x(\mathrm{g}) \rangle$.
For longer chains (i.e., $N=64$ and 96), the computation is very costly
in the lower-frequency regime~\cite{computeCost}. However, it was observed that, in this
regime, the averaging over even $100$ cycles is good enough to give a
smooth $x(\mathrm{g})$ vs $\mathrm{g}$ loop. To be on a safer side, we
have used $200$ cycles for averaging after leaving the first 20
cycles for the system to reach the stationary state. For the force 
amplitude $G$ and the frequencies $\omega$ used in this work, the average 
extension, $\langle x(\mathrm{g}) \rangle$, for the forward and the
backward paths is not the same and a hysteresis loop is observed. The
area of the hysteresis loop, $A_{\mathrm{loop}}$, defined as
\begin{equation}
    A_{\mathrm{loop}} = \oint \langle x (\mathrm{g}) \rangle
	d\mathrm{g},
\end{equation}
depends on the frequency $\omega$ and the amplitude $G$ of the
periodic force and serves as a dynamical order
parameter~\cite{chakrabarti_dynamic_1999}. The area of the loop,
$A_{\mathrm{loop}}$ is obtained numerically using the trapezoidal rule
after dividing the interval $\mathrm{g} \in [0,G]$ into $10^5$ equally
spaced intervals, and interpolating the value of $\langle x(\mathrm{g})
\rangle$ at the ends of these intervals using cubic splines of GNU
Scientific Library~\cite{galassi_GSL_2009}.

\section{Results and Discussions} \label{sec:results}

\subsection{Static case}

Let us first consider the equilibrium case where the dsDNA is subjected
to a constant pulling force, i.e., $\mathrm{g}(t) = G$ and check whether
the average separation behaves similarly as that obtained from the DSAW
model.

\figTwo

In Figs.~\ref{fig:2}(a) and ~\ref{fig:2}(b), we have plotted the
average separation between the strands of the DNA, where a constant
pulling force $G$ is acting, at various $G$ values for DNA having
$N=64$, 96, 128, 192, and 256 base pairs at temperatures $T=0.1$
and $T=0.2$, respectively. For smaller values of the force, the average
separation, $\langle x (G) \rangle$, which acts as an order parameter,
is zero showing that the two strands of the DNA are in the zipped phase.
On increasing the force value, the average separation abruptly increases
at some critical force value, $G_c(T)$, which depends on the
temperature, and $\langle x(G) \rangle \sim N$ showing that the DNA is
in the unzipped phase. The critical value of force $G_c(T)$, can be
obtained by using the finite-size scaling (FSS) of the form
\begin{equation}
\label{eq:fssx}
	\langle x (G) \rangle = N^{d} \mathcal{G} \left(  \frac{(G -
	G_c)}{N^{\phi}} \right),
\end{equation}
where $d$ and $\phi$ are the critical exponents. In
Figs.~\ref{fig:2}(c) and \ref{fig:2}(d), we
have plotted the scaled separation $\langle x (G) \rangle /N^d$ for the DNA of
various chain lengths as a function of
$(G-G_c)/N^{\phi}$ at $T=0.1$ and $T=0.2$, respectively. The
data for various chain lengths collapse on a scaling curve for the set
of critical exponents $d=1.10 \pm 0.05$, $\phi = 1.00 \pm 0.05$ with
critical force value $G_c(T=0.1) = 0.45 \pm 0.05$ for $T=0.1$, and exponents
$d=1.05 \pm 0.05$, $\phi = 1.00 \pm 0.05$ with $G_c(T=0.2)=0.25\pm0.05$
for $T=0.2$. The unzipping exponents $d=1$ and $\phi=1$, which are same
as the exponents obtained for the $D=1+1$ case,  show that the scaled
mean separation between the strands behave, in the thermodynamic limit
(i.e., $N \to \infty)$, as
\begin{equation}
	\langle x (G) \rangle/N \sim \left\{ \begin{array}{c c c} X_{-} &
	\text{for} & G < G_c(T) \cr X_{+} & \text{for} & G > G_c(T)  \end{array}
	\right. ,
\end{equation}
i.e., having two different values with a jump discontinuity at $G_c(T)$,
implying the first-order nature of the unzipping
transition~\cite{bhattacharjee_unzipping_2000,marenduzzo_phase_2001,marenduzzo_dynamical_2002,kapri_complete_2004}.

\figThree

In the absence of a pulling force, the free ends of the DNA can
move freely in both the $x$ and $y$ directions and perform more like
self-avoiding walks (SAWs). Let $(x_1, y_1)$ and $(x_2,y_2)$ represent
the coordinates of the end monomers of two strands of the DNA.  The
distance between the end monomers can then be obtained by $\mathrm{r} =
\sqrt{(x_2 - x_1)^2 + (y_2 - y_1)^2}$. In Fig.~\ref{fig:3}(a), we have
plotted the average separation $\langle \mathrm{r} \rangle$ as a
function of temperature $T$ for the DNA of various lengths $N=64$, 96,
128, 192, and 256. To estimate the melting temperature $T_M$ of the DNA, we use
the FSS of the form
\begin{equation} \label{eq:fssT}
	\langle \mathrm{r} \rangle \sim N^{d_t} \mathcal{Y} \left( ({T -
	T_M}) N^{\phi_t} \right),
\end{equation}
 where $d_t$ and $\phi_t$ are the critical exponents for the
 denaturation transition. When the scaled separation $\langle \mathrm{r}
 \rangle / N^{d_t}$ for various chain lengths are plotted as a function
 of $(T - T_M) N^{\phi_t}$, a nice collapse is obtained for the exponent
 values $d_t = 0.75 \pm 0.05$, $\phi_t = 1.00 \pm 0.05$ and $T_M = 0.30
 \pm 0.01$ [see Fig..~\ref{fig:3}(b)]. The melting of a dsDNA is a continuous transition in our model. Note that, at $T \ge T_M$, the
 exponent $d_t$ depends on dimensions as expected. In 2D, the value $d_t = 0.75$ 
 is consistent with the size exponent $\nu = 3/4$ of a SAW in
 2D~\cite{rubinstein2003}, whereas, in $D=1+1$, the end separation
 performs a random walk in 1D and the exponent $d_t =
 0.5$ (see, e.g., Ref. ~\cite{yadav_unzipping_2021}) is consistent with
 the size exponent $\nu = 1/2$ of a random
 walker~\cite{rubinstein2003}.

 Once the melting temperature $T_M$ for the model is obtained, it
 is easy to estimate the characteristic hydrogen bond energy
 $\varepsilon$ in real units and compare our results with the unzipping
 experiments. If $T_M^{*}$ represents the melting temperature in real
 units, then it is related to $T_M$ by $T_M = k_B T_M^{*}/\varepsilon$. Using
 $T_M=0.3$, and $T_M^{*}$ = 363 K~\cite{danilowicz_measurement_2004}, we
 obtain $\varepsilon \approx 0.1$eV.  Considering $\sigma=5.17\AA$ as the
 distance at which the interparticle  potential between the base pairs
 goes to zero, and $m=5\times10^{-22}$g as the average mass of each
 monomer, the unit of time is obtained as $\tau = \sqrt{m
 \sigma^2/\varepsilon} \approx 3$ps [see
 Ref.~\cite{mishra_dynamical_2013} and references therein]. The time and
 the distances are measured in real units as $t^{*} = \tau t$ and $r^{*}
 = \sigma r$, respectively. The order of the force is given by
 $\sigma/\AA \sim 160$pN. Using similar arguments as in
 Ref.~\cite{mishra_dynamical_2013}, the temperature conversion formula
 to real units below the melting temperature for our model can be
 obtained as $ T^{*} = 363 + 280(T - 0.30)$ K. Therefore, the reduced
 temperature $T = 0.1$ simulated in our paper corresponds to 307 K
 (i.e., $34^{\circ}$C). The critical force $G c = 0.45$ in reduced units
 corresponds to $G_c^{*} \approx 14$ pN similar to the critical force
 obtained in the experiments~\cite{danilowicz_measurement_2004}.

\subsection{Dynamic case}

From earlier studies, it is known that when a dsDNA is subjected to a
periodic force it can be unzipped either by keeping the amplitude $G$
fixed and changing the frequency $\omega$ or \textit{vice versa}. If $G$
is not very small, and $\omega$ is sufficiently high to avoid
equilibration of the DNA, then we obtain a hysteresis loop for the average
extension $\langle x (\mathrm{g}) \rangle$, whose area,
$A_{\mathrm{loop}}$, depends on $G$ and $\omega$. In
Fig.~\ref{fig:4}(a), we have shown the behavior of $A_{\mathrm{loop}}$
as a function of $\omega$ for the DNA of length $N=64$ for $G=1.5$ at
$T=0.1$. The area of the loop increases with the frequency, reaches a
maximum and then decreases as the frequency is increased further. The
loops at four different frequencies, labeled by $P,\ Q,\ R$, and $S$ in
Fig.~\ref{fig:4} (a), are shown in Fig.~\ref{fig:4}(b). Since the force
amplitude $G=1.5$ is about three times the critical force needed to
unzip the DNA at $T=0.1$, the stationary state of the DNA is unzipped
state. At a higher frequency $\omega_P = 6.28 \times 10^{-3}$, the
applied force fluctuates very rapidly and the DNA does not get time to
respond to this change. As a result, the DNA remains in the unzipped
state, as indicated by the higher values of the average extension,
$\langle x \rangle$, with a small loop area. On decreasing the frequency
to $\omega_Q = 6.28 \times 10^{-4}$, the DNA still remains in the
unzipped phase but with slightly increase in the loop area. On
decreasing the frequency further to $\omega_R = 6.28 \times 10^{-5}$,
the DNA gets enough time to relax to the oscillating force. Therefore,
during the portion of the cycle where the instantaneous force value is
less than the critical force $G_c$, the two strands of the DNA come
together and the complementary base pairs are formed, resulting the DNA
in the zipped phase with a large hysteresis loop area. This is indicated
by the lower values of the average extension, $\langle x \rangle$ for
smaller $\mathrm{g}$ values in Fig.~\ref{fig:4}(b). On decreasing the
frequency further to $\omega_S = 6.28 \times 10^{-6}$, the two strands
have ample time to relax in the lower as well as higher values of force
thus resulting in a very small loop area in the transition region. This
loop area will eventually go to zero on decreasing the frequency
further.   

\figFour

In Fig.~\ref{fig:5}, we have plotted $A_{\mathrm{loop}}$ as a function
of $\omega$ at various force amplitudes $G$ for the DNA of length
$N=64$. The figure shows that the frequency, $\omega^{*}(G)$, at which
the loop area is maximum depends on the amplitude $G$ of the oscillating
force. We observe that for smaller $G$ values, the
$A_{\mathrm{loop}}$ curves have broader peaks. The peak becomes narrower
with the increase in the force amplitude. Furthermore, on increasing
$G$, it is also observed that for amplitudes $G < 2G_c$, the position of
the peak [i.e., $\omega^{*}(G)$] increases toward higher frequencies,
whereas for values $G > 2G_c$, the peak position shifts toward lower
frequencies. It is not easy to give an exact cause for this behavior as
both $G$ and $\omega$ are competing with each other in this region.
Also, note that for $G < 2G_c$, the steady state of the DNA is a zipped
configuration and it cannot be fully unzipped.  Whereas, for $G > 2G_c$,
the steady state of the DNA is an unzipped configuration. Consequently,
the way the hysteresis loops are formed for the two cases are different
and have different shapes~\cite{kapri_unzipping_2014}. It is quite
plausible that the dependence of maximum $A_{\mathrm{loop}}$ on the
frequency might be completely different in these two different regions.
The figure also reveals that the height of the peak increases on
increasing $G$ value.  These observations are similar to the behavior
seen for the $A_{\mathrm{loop}}$ in MC simulations for the homopolymer
DNA~\cite{kapri_unzipping_2014}.  There is one striking feature,
the oscillatory behavior of $A_{\mathrm{loop}}$ at higher frequencies
for larger $G$ values, which was observed in MC simulations and
explained as higher Rouse modes~\cite{kapri_unzipping_2014}, is not
observed with the parameters used in this study. However, the presence
of such oscillatory behavior of $A_{\mathrm{loop}}$ has been reported in
LD simulations with different parameters~\cite{pal_dna_2018}. This needs
further exploration.  

\figFive

\figSix

The loop area, $A_{\mathrm{loop}}$ as a function of frequency $\omega$
for amplitude $G=1$ and $3$, are plotted in Figs.~\ref{fig:6}(a) and
\ref{fig:6}(b), respectively, for the DNA of various chain lengths
$N=16$, 32, 64 and 96. The figure shows that, similar to the MC
simulation studies~\cite{kapri_unzipping_2014}, the peak of the area
curves shift toward the lower-frequency side on increasing the chain
length. Furthermore, these plots also show that the maximum of the loop
area increases with amplitude $G$. We use FSS of the form    
\begin{equation} \label{eq:ANscale}
    A_{\mathrm{loop}} = N^{d} \mathcal{A} \left( \omega N^z \right),
\end{equation}
to obtain the behavior of $A_{\mathrm{loop}}$ in the thermodynamic limit
from finite-size chains. We obtain a nice collapse for exponents $d=1.00
\pm 0.05$ and $z=1.00 \pm 0.05$ for $G=1$ [Fig.~\ref{fig:6}(c)]. These
exponent values are same as that obtained in MC simulation
study~\cite{kapri_unzipping_2014}. However, for higher force amplitudes
(e.g., $G=3$) we get a reasonable collapse for a slightly higher value
$d=1.15 \pm 0.05$. Figures~\ref{fig:6}(c) and \ref{fig:6}(d) show that there are
strong finite-size effects and the curves for the smallest chain length
$N=16$ considered in this study do not collapse perfectly on the scaling
curve. However, the data for the higher chain lengths, $N=64$ and 96,
collapse perfectly on the scaling curve for $G=1$. In order to 
improve the quality of data collapse at force amplitude $G = 3$, longer 
chain lengths need to be simulated. But, due to the higher computation 
cost~\cite{computeCost}, these simulations were not performed. The
exponents $d=1$ and $z=1$ show that the loop area scales as
$A_{\mathrm{loop}} \sim 1/\omega$ in the high-frequency regime.

\figSeven

To obtain the behavior of $A_{\mathrm{loop}}$ at lower-frequency regime,
we have plotted in Fig.~\ref{fig:7}(a) (in a semilog scale) the loop
area as a function of $\omega^{\beta}(G-G_c)^{\alpha}$, where $G_c$ is
the critical force for the static force case, obtained using LD
simulations for the DNA of length $N=64$ at various force amplitudes
$G=1.0$, 1.25, 1.5, 2.0, and 3.0 at temperature $T=0.1$. We obtain an
excellent data collapse for exponents $\alpha = 1.08 \pm 0.03$
and $\beta = 1.25 \pm 0.03$ and critical force $G_c(T=0.1)=0.45$
obtained for the static force case in the previous section
(Eq.~\eqref{eq:fssx}). The exponents and the errors in them are
estimated by minimizing the variance obtained from $A_{loop}$ curves for
various $G$ values integrated over a decade in
frequency~\cite{newman1999}. In Fig.~\ref{fig:7}(b), the collapse
obtained for $A_{\mathrm{loop}}$ curves for chain length $N=32$ at
$T=0.2$ for three different force amplitudes $G=1.0$, $1.5$, and $2.0$
are plotted with $G_c(T=0.2) = 0.25$ and $\alpha = 1.09\pm0.03$ and
$\beta=1.24\pm0.02$. The quality of the collapse indicates that the
exponent values $\alpha$ and $\beta$ are independent of temperature
used. Furthermore, these exponent values are similar to that obtained
in previous studies using MC simulations of a DSAW model of the
homopolymer DNA at $T=1$~\cite{kapri_unzipping_2014}, and the double
stranded block copolymer DNA at $T=4$~\cite{yadav_unzipping_2021}. In
the homopolymer DNA study, the $A_{\mathrm{loop}}$ was plotted against
$G^{\alpha}\omega^{\beta}$. To check the quality of collapse with newer
scaled function, we have plotted the $A_{\mathrm{loop}}$ data, for the
chain length $N=512$, obtained in Ref.~\cite{kapri_unzipping_2014} as a
function of $\omega^{\beta}(G-G_c)^{\alpha}$ with $G_c (T=1) =0.678$ and
$\alpha=0.99\pm0.03$ and $\beta=1.25\pm0.03$. The quality of the
plot shown in Fig.~\ref{fig:7}(c) is found to be even better than the
plot shown with function $G^{\alpha}\omega^{\beta}$ in
Ref.~\cite{kapri_unzipping_2014}.

We can use the behavior of $A_{\mathrm{loop}}$ in low- and high-frequency
regimes to obtain the scaling function $\mathcal{G}(\omega)$. At lower
frequencies (i.e., $\omega \to 0$), we observed that, for large $N$, the
$A_{\mathrm{loop}}$ scales as $G^{\alpha} \omega^{\beta}$, while at
higher frequencies (i.e., $\omega \to \infty$), $A_{\mathrm{loop}} \sim
1/\omega$ [from Eq.~\eqref{eq:ANscale}]. These requirements are
satisfied by the scaling function  
\begin{equation} \label{eq:scalingCurve}
	\mathcal{G}(\omega) = \frac{B G^{\alpha} \omega^{\beta} }{ \omega^{d +
	\beta} + C^2}, 
\end{equation}
with $B$ and $C$ as the fitting parameters. The scaling function,
$\mathcal{G}(\omega)$, for $G=1$ with exponent $d=1$, and parameters
$B=0.01$ and $C=0.005$ obtained by data fitting, is plotted in
Fig.~\ref{fig:6}(c) by a solid line. The function fits the data
extremely well in the frequency range extended over more than four
decades. We have also plotted the same scaling function for $G=3$, with
exponent $d=1.15$ for parameters $B=0.001$ and $C=0.002$ in
Fig.~\ref{fig:6}(d). Although the scaling function $\mathcal{G}(\omega)$
fits reasonably well in the lower-frequency regime, it however deviates
with the scaled data at higher frequencies. From MC simulation
studies~\cite{kapri_unzipping_2014}, we know that for $G=3$,
$A_{\mathrm{loop}}$ exhibits an oscillatory behavior in the
higher-frequency regime (visible only for longer chain lengths) and the
above scaling form is not suitable. Since the maximum chain length simulated
in this study, is still 5 times lesser than that simulated in
Ref.~\cite{kapri_unzipping_2014}, we do not see the oscillatory behavior
of $A_{\mathrm{loop}}$ at higher frequencies. The deviation of the
simulation data from the scaled curve $\mathcal{G}(\omega)$ for $G=3$ at
higher-frequency regime in Fig.~\ref{fig:7}(d) may be due to this very
reason.

\section{Conclusions}\label{sec:summary}

We study the unzipping of a dsDNA subjected to a periodic force with
amplitude $G$ and frequency $\omega$ using extensive LD simulations on
longer DNA chains, having up to 192 monomers with $N=96$ base
pairs, that are six times longer than previous LD simulation
studies~\cite{kumar_statistical_2013,mishra_dynamical_2013,mishra_scaling_2013,kumar_periodically_2016,pal_dna_2018}.
We first study the static force case and obtain the equilibrium average
separation between the strands of the DNA, $\langle x (\mathrm{G})
\rangle$, as a function of force $\mathrm{G}$ at two different
temperatures ($T=0.1$ and $T=0.2$). Using the FSS of force-distance
isotherms $\langle x (\mathrm{G}) \rangle$ for various chain lengths
$N=64$, 96, 128, 192, and 256, we obtain the dimensionless
critical force $G_c(T=0.1) = 0.45 \pm 0.05$ at $T=0.1$ and $G_c(T=0.2) =
0.25 \pm 0.05$ at $T=0.2$, needed to unzip the DNA in the thermodynamic
limit $N \to \infty$. The FSS reveals that the scaled average separation
between the strands, $\langle x \rangle/N$, has a jump discontinuity at
$G_c$ implying a first-order nature of the phase transition similar to
the DSAW model studied
earlier~\cite{bhattacharjee_unzipping_2000,marenduzzo_phase_2001,marenduzzo_dynamical_2002,kapri_complete_2004}.
We also obtained the melting temperature $T_M=0.30 \pm 0.01$ for
the model. The melting of DNA is a continuous transition in our model.
When the DNA is subjected to a periodic force, the average separation
between the strands $\langle x (\mathrm{g}) \rangle$, when plotted
against $\mathrm{g}$, shows hysteresis whose area, $A_{\mathrm{loop}}$,
depends on the amplitude $G$ and the frequency $\omega$ of the
oscillating force. On decreasing the frequency, the loop area first
increases from zero, reaches a maximum value at some frequency
$\omega^{*}(G)$, which depends on the amplitude $G$, and then decreases
to zero again at lower frequencies. The FFS scaling of
Eq.~\eqref{eq:ANscale} shows that, in the thermodynamic limit, the loop
area scales as $A_{\mathrm{loop}} \sim 1/\omega$ in the higher-frequency
regime. In contrast, the loop area which scales as $A_{\mathrm{loop}}
\sim (G-G_c)^{\alpha}\omega^{\beta}$ is found to have exponent values
$\alpha\approx1$ and $\beta\approx1.25$, The exponent values,
which are found to be temperature independent, are same as that obtained in
earlier unzipping studies of homopolymer DNA~\cite{kapri_unzipping_2014}
and block copolymer DNA~\cite{yadav_unzipping_2021} by a periodic force
on a DSAW model using MC simulations at two different temperatures. The
fact that we have obtained the same values for the exponents, $\alpha$
and $\beta$, at various temperatures for two different problems, i.e.,
homopolymer DNA and block copolymer DNA, where the former is studied by
two different methods, MC and LD simulations of longer chain lengths at
different temperatures strongly indicates that $\alpha=1.0$ and
$\beta=1.25$ are the true scaling exponents for the DNA unzipping
problem that quantify the decrease of $A_{loop}$ to zero at low
frequencies at all temperatures. Single molecule manipulation
experiments can shed more light on these scaling exponents.          

\section*{Acknowledgements}

I thank S. M. Bhattacharjee, A. Chaudhuri and R. Yadav for their
comments on the manuscript. I thank A. Chaudhuri for allowing me to use
his computational resources.

\end{document}